
\input harvmac.tex
\Title{CTP/TAMU-80/92}{{Classical Dynamics of Macroscopic Strings}
\footnote{$^\dagger$}{Work supported in part by NSF grant PHY-9106593.}}

\centerline{
Ramzi~R.~ Khuri\footnote{$^*$}{Supported by a World Laboratory Fellowship.}}
\bigskip\centerline{Center for Theoretical Physics}
\centerline{Texas A\&M University}\centerline{College Station, TX 77843}

\vskip .3in
In recent work, Dabholkar {\it et al.} constructed static ``cosmic string"
solutions of the low-energy supergravity equations of the heterotic string,
and conjectured that these solitons are actually
exterior solutions for infinitely long fundamental strings. In this
paper we provide compelling dynamical evidence to support this conjecture
by computing the dynamical force exerted by a solitonic string on an
identical test-string limit, the Veneziano amplitude for the scattering of
macroscopic winding states and the metric on moduli space for the scattering
of two string solitons. All three methods yield trivial scattering in the
low-energy limit.

\Date{12/92}
\def\x{{(R^2-2)v^2\over 1-v^2}}

\lref\quartet{D.~J.~Gross,
J.~A.~Harvey, E.~J.~Martinec and R.~Rohm, Nucl. Phys. {\bf B256} (1985) 253.}

\lref\dine{M.~Dine, Lectures delivered at
TASI 1988, Brown University (1988) 653.}

\lref\bpst{A.~A.~Belavin, A.~M.~Polyakov, A.~S.~Schwartz and Yu.~S.~Tyupkin,
Phys. Lett. {\bf B59} (1975) 85.}

\lref\rkinst{R.~R.~Khuri, Phys. Lett. {\bf B259} (1991) 261.}

\lref\ccrk{C.~G.~Callan and R.~R.~Khuri, Phys. Lett. {\bf B261} (1991) 363.}

\lref\rkmant{R.~R.~Khuri, Nucl. Phys. {\bf B376} (1992) 350.}

\lref\rkthes{R.~R.~Khuri, {\it Solitons and Instantons in String Theory},
 Princeton University Doctoral Thesis, August 1991.}

\lref\mono{R.~R.~Khuri, Phys. Lett. {\bf B294} (1992) 325.}

\lref\monscat{R.~R.~Khuri, Phys. Lett. {\bf B294} (1992) 331.}

\lref\rkgeo{R.~R.~Khuri, ``Geodesic Scattering of Solitonic Strings",
Texas A\&M preprint CTP/TAMU-79/92.}

\lref\dghrr{A.~Dabholkar, G.~Gibbons, J.~A.~Harvey and F.~Ruiz Ruiz,
Nucl. Phys. {\bf B340} (1990) 33.}

\lref\dabhar{A.~Dabholkar and J.~A.~Harvey,
Phys. Rev. Lett. {\bf 63} (1989) 478.}

\lref\prso{M.~K.~Prasad and C.~M.~Sommerfield, Phys. Rev. Lett. {\bf 35}
(1975) 760.}

\lref\mantone{N.~S.~Manton, Nucl. Phys. {\bf B126} (1977) 525.}

\lref\manttwo{N.~S.~Manton, Phys. Lett. {\bf B110} (1982) 54.}

\lref\mantthree{N.~S.~Manton, Phys. Lett. {\bf B154} (1985) 397.}

\lref\atiyah{M.~F.~Atiyah and N.~J.~Hitchin, Phys. Lett. {\bf A107}
(1985) 21.}

\lref\atiyahbook{M.~F.~Atiyah and N.~J.~Hitchin, {\it The Geometry and
Dynamics of Magnetic Monopoles}, Princeton University Press, 1988.}

\lref\gsw{M.~B.~Green, J.~H.~Schwartz and E.~Witten,
{\it Superstring Theory} vol. 1, Cambridge University Press (1987).}

\lref\polch{J.~Polchinski, Phys. Lett. {\bf B209} (1988) 252.}

\lref\dfluone{M.~J.~Duff and J.~X.~Lu, Nucl. Phys. {\bf B354} (1991) 141.}

\lref\dflutwo{M.~J.~Duff and J.~X.~Lu, Nucl. Phys. {\bf B354} (1991) 129.}

\lref\dfluthree{M.~J.~Duff and J.~X.~Lu, Phys. Rev. Lett. {\bf 66}
(1991) 1402.}

\lref\dflufour{M.~J.~Duff and J.~X.~Lu, Nucl. Phys. {\bf B357} (1991)
534.}

\lref\dfstel{M.~J.~Duff and K.~S.~Stelle, Phys. Lett. {\bf B253} (1991)
113.}

\lref\gh{G.~W.~Gibbons and S.~W.~Hawking, Phys. Rev. {\bf D15}
(1977) 2752.}

\lref\ghp{G.~W.~Gibbons, S.~W.~Hawking and M.~J.~Perry, Nucl. Phys. {\bf B318}
(1978) 141.}

\lref\briho{D.~Brill and G.~T.~Horowitz, Phys. Lett. {\bf B262} (1991)
437.}

\lref\raj{R.~Rajaraman, {\it Solitons and Instantons}, North Holland,
1982.}

\lref\bogo{E.~B.~Bogomolnyi, Sov. J. Nucl. Phys. {\bf 24} (1976) 449.}

\lref\dflufive{M.~J.~Duff and J.~X.~Lu, Class. Quant. Grav. {\bf 9}
(1992) 1.}

\lref\dflusix{M.~J.~Duff and J.~X.~Lu, Phys. Lett. {\bf B273} (1991)
409.}

\lref\chsone{C.~G.~Callan, J.~A.~Harvey and A.~Strominger, Nucl. Phys.
{\bf B359} (1991) 611.}

\lref\chstwo{C.~G.~Callan, J.~A.~Harvey and A.~Strominger, Nucl. Phys.
{\bf B367} (1991) 60.}

\lref\strom{A.~Strominger, Nucl. Phys. {\bf B343} (1990) 167.}

\lref\duff{M.~J.~Duff, Class. Quant. Grav. {\bf 5} (1988).}

\lref\felce{A.~G.~Felce and T.~M.~Samols, ``Low-Energy Dynamics of String
Solitons" NSF-ITP-92-155, November 1992.}

\lref\cfmp{C.~G.~Callan, D.~Friedan, E.~J.~Martinec
and M.~J.~Perry, Nucl. Phys. {\bf B262} (1985) 593.}

\lref\ckp{C.~G.~Callan,
I.~R.~Klebanov and M.~J.~Perry, Nucl. Phys. {\bf B278} (1986) 78.}

\newsec{Introduction}

Soliton solutions of string theory recently discovered by Dabholkar {\it et
al.} \refs{\dghrr,\dabhar} have the property, like BPS magnetic
monopoles, that they exert zero static force on each other and can be
superposed to form multi-soliton solutions with arbitrarily variable
collective coordinates. In this paper we show that, in
contradistinction to the BPS case, the velocity-dependent forces
between these string solitons also vanish (i.e. we argue that the
scattering is trivial).  We also
argue that this phenomenon provides further, dynamical evidence for the
identification of the Dabholkar-Harvey soliton with the underlying
fundamental string by comparing the scattering of these soliton
solutions with expectations from a Veneziano amplitude computation for
macroscopic fundamental strings.

In section 2, we first summarize the solution of
Dabholkar {\it et al.}\dghrr, who construct static ``cosmic string''
solutions of the low-energy supergravity equations of the heterotic
string.  These solutions have several remarkable properties.  The most
notable is the vanishing of the static force between parallel straight
``cosmic'' strings of like orientation.  This feature is the result of
a cancellation of the long-range forces due to exchange of axions,
gravitons and dilatons, and is reminiscent of the cancellation of gauge
and Higgs forces between BPS monopoles. Indeed, in perfect analogy with
the BPS case, the no-force condition makes it possible to construct
multi-soliton solutions with any number of parallel, like-oriented
straight cosmic strings at arbitrary separations. Since these static
properties are also possessed by fundamental strings winding around an
infinitely large compactified dimension, Dabholkar {\it et al.}
conjecture \dabhar~ that the soliton is actually the exterior solution
for an infinitely long fundamental string.

We examine the scattering of these solitons in section 3
using the ``test string'' approximation.  From the sigma model
action describing the motion of a point-like test string
in a general background of axion, graviton and dilaton fields, we
derive an effective action for the motion of the center of mass
coordinate of the test string in the special background provided by a
string soliton. We of course find that the static force vanishes (this
is how Dabholkar {\it et al.} constructed their ansatz for the string
soliton). More remarkably, we find that the $O(v^2)$ velocity-dependent
forces vanish as well. This result suggests that there is trivial
scattering between these string solitons, at least in the test-string
approximation.

We address the scattering problem in section 4 from the string
theoretic point of view. In particular, we calculate the string
four-point amplitude for the scattering of macroscopic winding state
strings in the infinite winding radius limit. In this scenario, we can
best approximate the soliton scattering problem considered in section 3.
We find that the Veneziano amplitude obtained also indicates trivial
scattering in the large winding radius limit, thus providing evidence
for the identification of the string soliton solutions with infinitely
long macroscopic fundamental strings.

In section 5 we compute the metric on moduli space for the string
soliton in $D=4$ to lowest nontrivial order in the string tension.
The geodesics of this metric represent the motion of quasi-static solutions
in the static solution manifold and in the absence of a full time dependent
solution provide a good approximation to the low-energy dynamics of the
solitons. The metric is found to be flat, which again implies trivial
scattering
of the solitons, in agreement with the results of the previous two sections.

We conclude in section 6 with a discussion of our results.
In particular, our findings provide compelling dynamical
evidence for the identification of the solitonic string with the underlying
fundamental string. We note that the role of these solitons in string theory
parallels the role of soliton solutions in field theory in describing extended
particle states.

The results of sections 3 and 4 have been previously summarized in
\ccrk. The Manton scattering result of section 5 has recently been
summarized in \rkgeo.

\newsec{String Multi-Soliton Static Solution}

In recent work\refs{\dabhar,\dghrr}, Dabholkar and others
presented a low-energy analysis of macroscopic superstrings and
discovered several interesting analogies between macroscopic superstrings
and solitons in supersymmetric field theories. The main result of this
work centers on the existence of exact multi-string solutions of the
low-energy supergravity super-Yang-Mills equations of motion. In
addition, Dabholkar {\it et al.} find a Bogomolnyi bound for the energy
per unit length which is saturated by these solutions, just as the
Bogomolnyi bound is saturated by magnetic monopole solutions in ordinary
Yang-Mills field theory.

The Dabholkar {\it et al.} solution may be outlined as follows.
The action for the massless spacetime fields
(graviton, axion and dilaton) in the presence of a source
string can be written as\refs{\cfmp,\ckp,\dghrr}
\eqn\sourceaction {S = {1\over 2\kappa^2}\int d^Dx\sqrt{g}\left(R-{1\over 2}
{{(\partial\phi)}^2}-{1\over 12}{e^{-2\alpha\phi}}H^2\right)+S_\sigma,}
with the source terms contained in the sigma model action $S_\sigma$ given by
\eqn\ssigma {S_\sigma=-{\mu\over 2}\int d^2\sigma(\sqrt{\gamma} \gamma^{mn}
\partial_mX^\mu\partial_nX^\nu g_{\mu\nu}e^{\alpha\phi} +
\epsilon^{mn}\partial_mX^\mu\partial_nX^\nu B_{\mu\nu}),}
with $\alpha=\sqrt{2/(D-2)}$ and $\gamma_{mn}$ a worldsheet metric to be
determined.

The sigma model action $S_\sigma$ describes the coupling of the string to the
metric, antisymmetric tensor field and dilaton. The first part of the
action $S$ above represents the effective action for the
massless fields in the spacetime frame and
whose equations of motion are equivalent to conformal invariance of the
underlying sigma model. The combined action thus generates the equations
of motion satisfied by the massless fields in the presence of a
macroscopic string source\dghrr.

The static solution to the equations of motion is given by \dghrr\
\eqn\ansatz{\eqalign{ds^2&=e^{A}[-dt^2+(dx^1)^2]+e^{B}d\vec x \cdot d\vec x\cr
A&={D-4\over D-2}E(r)\qquad B=-{2\over D-2}E(r)\cr
\phi&=\alpha E(r)\qquad\qquad B_{01}=-e^{E(r)},\cr}}
where $x^1$ is the direction along the string, $r=\sqrt{\vec x \cdot\vec x}$
and
\eqn\ssoln{e^{-E(r)}=\cases{1+{M\over r^{D-4}}&$D>4$\cr 1-8G\mu\ln(r)&$D=4$
\cr}}
for a single static string source. The solution can be generalized to
an arbitrary number of static string sources by linear superposition of
solutions of the ($D-2$)-dimensional Laplace's equation.

The force exerted on a test string moving in given background fields
is obtained from the sigma model equation of motion\dghrr
\eqn\euler {\nabla_m(\gamma^{mn}\nabla_nX^\mu)=-\Gamma^\mu_{\nu\rho}
\partial_mX^\nu\partial_nX^\rho\gamma^{mn}+\half
H^\mu{}_{\nu\rho}\partial_mX^\nu\partial_nX^\rho\epsilon^{mn},}
where $\Gamma^\mu_{\nu\rho}$ are the Christoffel symbols calculated from
the sigma model metric $G_{\mu\nu}=g_{\mu\nu}e^{\alpha\phi}$. We make the
usual distinction between the sigma model metric and the Einstein metric;
spacetime indices are raised and lowered by contraction with $G_{\mu\nu}$;
worldsheet indices are denoted by $m$ and $n$.

We consider a stationary test string in the background of a source
string located at the origin. Assume further that both strings run along
the $x_1$ direction and have the same orientation. We use conformal
gauge for the test string and get $X^0=\tau$, $X^1=\sigma$,
$\gamma_{mn}=diag(-1,+1)$ and $\epsilon^{01}=+1$. The transverse force
then vanishes\dghrr:
\eqn\statzero{ {d^2\over d\tau^2} X^i=-2\Gamma^i_{00}+H^i{}_{10}=0.}
Note that if the test string and source string were oppositely oriented
then the second term would appear with a negative sign and there would
be a net attractive force. Also note that the no-force condition
depends only on the general ansatz \ansatz\ and not on the precise form
of the solution in \ssoln.

The zero-force condition arises from the cancellation of
long-range forces of exchange of the massless fields of the string
(the graviton, axion and dilaton) and can be seen explicitly from
\statzero. This is a perfect analog to the zero-force condition of
Manton for magnetic monopoles, which requires that the attractive
scalar exchange force precisely cancel the repulsive vector exchange
force when the Bogomolnyi bound is attained. Dabholkar {\it et al.}\dghrr\
show that a similar Bogomolnyi bound is satisfied by their string
soliton solutions, further strengthening the analogy with the
monopoles. In the next section, we use the test-string approach
to study the dynamics of these solitons.

\newsec{Test String Approximation}

We now turn to the dynamics of these string solitons. While the static
force vanishes as a result of the cancellation of long-range forces of
exchange, the force between two moving solitons is in general
nonvanishing and depends on the velocities of the solitons. The most
complete answer would be given by a full time-dependent solution of the
equations of motion of the above action for the case of an arbitrary
number of sources moving with arbitrary transverse velocities. These
equations, however, are much more difficult to solve for moving sources
than for a static configuration. Even a two-soliton
solution is in general quite intractable for this class of actions. The
next best step is the calculation of the Manton metric on moduli
space\manttwo, and this will be shown in section 5.

Here we will take the preliminary ``test-string" approach: we consider
simply a test string moving in the background of a string soliton whose
massless fields are given by \ansatz. The advantage of this approach is
that we obtain a first order approximation in a relatively simple
manner. The effective Lagrangian for the motion of a test string in the
background of the source string can be read off from the sigma model
source action $S_\sigma$. We then solve the constraint equation for the
worldsheet metric obtained by varying the worldsheet Lagrangian ${\cal
L}$. The resultant solution for the worldsheet metric along with the
static solution for the spacetime metric, antisymmetric tensor field
and dilaton from the static ansatz for a single source string are then
substituted into the Lagrangian, whose equations yield the dynamics of
the test string in the source string background.

The constraint equation for the worldsheet metric is given by
\eqn\constraint {F_{mn}-{1\over
2}\gamma_{mn}\left(\gamma^{ab}F_{ab}\right)=0,}
where $F_{mn}\equiv \partial_m X^\mu \partial_n X^\nu G_{\mu\nu}$.
The solution to \constraint\ is given by
\eqn\worldmetric {\gamma_{mn}=\Omega(X^\rho)\partial_m X^\mu\partial_n X^\nu
G_{\mu\nu},}
where $\Omega(X^\rho)$ is an arbitrary conformal factor. Note that the choice
$\Omega=e^{-E}$ fixes the conformal gauge
$\gamma^{mn}=\eta^{mn}$ in the static case. Substituting \worldmetric\
and \ansatz\ in the worldsheet action gives
\eqn\worldsheet {{\cal L}=-\mu\left[\sqrt{-\det\left(\eta_{mn}e^E
+\partial_m X^\mu\partial_n X^\nu\delta_{\mu\nu}\right)}-e^E\right].}
Naturally $\Omega$ drops out of $\cal L$. The relative sign of the two terms in
\worldsheet\ would have been ``plus'' for oppositely oriented source
and test strings. Taking the transverse
coordinates of the test string independent of $x^1$, the action reduces to
\eqn\flatlagrange{\eqalign {{\cal L}&=-\mu\left[e^E\left(1-{(\dot X^i)^2\over
e^E}\right)^{1\over 2}-e^E\right]~~.\cr}}
Expanding this in powers of velocity one easily obtains
\eqn\lowv{{\cal L} = {\mu\over 2}(\dot X^i)^2 +O(\dot X^4)~~.}
So, the same ansatz which causes the static force on a test string to vanish,
causes the lowest-order velocity-dependent force to vanish. In this
approximation, then, we have trivial scattering (i.e. no deviation from
initial trajectories) and the Manton metric on moduli space is
flat. Note that nothing in this argument
depended on the detailed form of $E$, so the same result holds for a string
moving in a general multi-string background.

Needless to say, this is a rather surprising result and so we would like
to obtain some kind of confirmation for this answer. We proceed to do
so in the next two sections. In section 4 we consider the scattering
problem from the purely string theoretic viewpoint, while in section 5
we explicitly compute the metric on moduli space in $D=4$ to lowest
nontrivial order in the string tension.

\newsec{Veneziano Amplitude for Macroscopic Fundamental Strings}

As mentioned earlier, it seems likely that the string solitons of Dabholkar
{\it et al.} are to be identified with infinitely long fundamental strings
of the underlying string theory. If that is so, the results we have just
found should agree with the corresponding Veneziano amplitude calculation of
string scattering. We now turn to this useful sanity check.

The scattering problem is set up in four dimensions, as the kinematics
correspond essentially to a four dimensional scattering problem, and
strings in higher dimensions generically miss each other anyway\polch.
The precise compactification scheme is irrelevant to our purposes.

The winding state strings then reside in four spacetime dimensions
$(0123)$, with one of the dimensions, say $x_3$, taken to be periodic with
period $R$, called the winding radius. The winding number $n$ describes
the number of times the string wraps around the winding dimension:
\eqn\wind{x_3\equiv x_3 + 2\pi Rn,}
and the length of the string is given by $L=nR$. The integer $m$, called
the momentum number of the winding configuration, labels the allowed
momentum eigenvalues. The momentum in the winding direction is thus
given by
\eqn\pthree{p^3={m\over R}.}
The number $m$ is restricted to be an integer so that the quantum wave function
$e^{ip\cdot x}$ is single valued.
The total momentum of  each string can be written as the sum of a
right momentum and a left momentum \eqn\tmoment{p^\mu=p^\mu_R+p^\mu_L,}
where $p^\mu_{R,L}=(E,E\vec v,{m\over 2R}\pm nR)$,
$\vec v$ is the transverse velocity and $R$ is the winding radius.
The mode expansion of the general
configuration $X(\sigma,\tau)$ in the winding direction
satisfying the two-dimensional wave equation
and the closed string boundary conditions can be written as the sum of
right moving pieces and left moving pieces, each with the mode expansion
of an open string\gsw~
\eqn\movers{\eqalign{X(\sigma,\tau)&=X_R(\tau -\sigma) + X_L(\tau +\sigma)\cr
X_R(\tau -\sigma)&=x_R + p_R(\tau -\sigma) + {i\over 2}
\sum_{n=0} {1\over n}\alpha_n e^{-2in(\tau - \sigma)}\cr
X_L(\tau +\sigma)&=x_L + p_L(\tau +\sigma) + {i\over 2}
\sum_{n=0} {1\over n}\tilde\alpha_n e^{-2in(\tau + \sigma)}.\cr}}
The right moving and left moving components are then essentially
independent parts with corresponding vertex operators, number operators
and Virasoro conditions.

The winding configuration described by $X(\sigma,\tau)$ describes a
soliton string state. It is therefore a natural choice for us to compare
the dynamics of these states with the solitons solutions of the previous
sections in order to determine whether we can identify the solutions of
the supergravity field equations with infinitely long fundamental
strings. Accordingly, we study the scattering of the winding states in
the limit of large winding radius.

Our kinematic setup is as follows. We consider the scattering of
two straight macroscopic strings in the CM frame with
winding number $n$ and momentum number $\pm m$ \refs{\gsw,\polch}.
The incoming momenta in the CM frame are given by
\eqn\imoment{\eqalign{p^\mu_{1R,L}&=(E,E\vec v,{m\over 2R}\pm nR)\cr
p^\mu_{2R,L}&=(E,-E\vec v,-{m\over 2R}\pm nR).\cr}}
Let $\pm m'$ be the outgoing momentum number.
For the case of $m=m'$, the outgoing momenta are given by
\eqn\omoment{\eqalign{-p^\mu_{3R,L}&=(E,E\vec w,{m\over 2R}\pm nR)\cr
-p^\mu_{4R,L}&=(E,-E\vec w,-{m\over 2R}\pm nR),\cr}}
where conservation of momentum and winding number have been used and
where $\pm\vec v$ and $\pm\vec w$ are the incoming and outgoing velocities of
the strings in the transverse $x-y$ plane. The outgoing momenta
winding numbers are not {\it a priori} equal to the initial winding
numbers, but must add up to $2n$. Conservation of energy for
sufficiently large $R$ then results in the above answer. This is also in
keeping with the soliton scattering nature of the problem (i.e. the
solitons do not change ``shape" during a collision).

For now we have assumed no longitudinal excitation ($m=m'$).
We will later relax this condition to allow for such excitation, but
show that our answer for the scattering is unaffected by this
possibility. It follows  from this condition that
$v^2=w^2$. For simplicity we take $\vec v=v\hat x$ and
$\vec w=v(\cos\theta\hat x+\sin\theta\hat y)$, and thus reduce the
problem to a two-dimensional scattering problem.

As usual, the Virasoro conditions $L_0=\widetilde{L}_0=1$ must hold, where
\eqn\vops{\eqalign{L_0&=N+\half (p^\mu_R)^2\cr \widetilde{L}_0&=\widetilde
{N}+\half (p^\mu_L)^2 \cr}}
are the Virasoro operators\gsw\ and where $N$ and $\widetilde{N}$ are the
number operators for the right- and left-moving modes respectively:
\eqn\numbs{\eqalign{N&=\sum \alpha^\mu_{-n}\alpha_{n\mu}\cr
\widetilde{N}&=\sum \tilde\alpha^\mu_{-n}\tilde\alpha_{n\mu},\cr}}
where we sum over all dimensions, including the compactified ones.
It follows from the Virasoro conditions that
\eqn\evr{\eqalign{\widetilde{N}-N&=mn\cr
	E^2(1-v^2)&=2N-2+{({m\over 2R}+nR)}^2.\cr}}

In the following we set $n=1$ and consider for simplicity the scattering
of tachyonic winding states. For our purposes, the nature of the string
winding states considered is irrelevant. A similar calculation for
massless bosonic strings or heterotic strings, for example, will be
slightly more complicated, but will nevertheless exhibit the same essential
behaviour. For tachyonic winding states we have $N=\widetilde{N}=m=0$.
Equation \evr\ reduces to
\eqn\tevr{E^2(1-v^2)=R^2-2.}
The Mandelstam variables $(s,t,u)$ are identical for right and left
movers and are given by
\eqn\mandlestam{\eqalign{s&=4\left[\x-2\right]\cr
t&=-2\left[\x\right](1+\cos\theta)\cr
u&=-2\left[\x\right](1-\cos\theta).\cr}}
It is easy to see that
$p_{iR}\cdot p_{jR}=p_{iL}\cdot p_{jL}$ holds
for this configuration so that the tree level 4-point function
reduces to the usual Veneziano amplitude for closed tachyonic strings\polch
\eqn\veneziano{\eqalign{A_4&={\kappa^2\over 4} B(-1-s/2,-1-t/2,-1-u/2)\cr
&=({\kappa^2\over 4}) {\Gamma(-1-s/2)\Gamma(-1-t/2)\Gamma(-1-u/2)\over
\Gamma(2+s/2)\Gamma(2+t/2)\Gamma(2+u/2)}.\cr}}
This can be seen as follows. In the standard computation of the four
point function for closed string tachyons, we rely on the independence
of the right and left moving open strings. For the tachyonic winding
state, we also separate the right and left movers with vertex operators
given by $V_R=e^{ip_R\cdot x_R}$ and $V_L=e^{ip_L\cdot x_L}$ respectively
to arrive at the following expression for the amplitude
\eqn\afour{A_4={\kappa^2\over 4}\int d\mu_4(z)\prod_{i<j}
|z_i-z_j|^{p_{iR}\cdot p_{jR}} |z_i-z_j|^{p_{iL}\cdot p_{jL}}.}
{}From $p_{iR}\cdot p_{jR}=p_{iL}\cdot p_{jL}$, \afour\ reduces to the
expression for the four-point amplitude of a nonwinding closed tachyonic
string, from which the standard Veneziano amplitude in \veneziano\ results.

To compare the implications of $A_4$ with the classical scattering of
section three we take $R\to\infty$. It is
convenient to define $x\equiv\x=s/4+2$, since the Mandelstam variables can
be expressed solely in terms of $x$ and $\theta$. We now have
$A_4=A_4(x,\theta)$, which can be explicitly written as
\eqn\ampone{A_4=({\kappa^2\over 4})
{\Gamma(3-2x)\Gamma(-1+x(1+\cos\theta))\Gamma(-1+x(1-\cos\theta))\over
\Gamma(-2+2x)\Gamma(2-x(1+\cos\theta))\Gamma(2-x(1-\cos\theta))}.}
The problem reduces to studying $A_4$ in the limit $x\to\infty$.
We now use the identity $\Gamma(1-a)\Gamma(a)\sin\pi a=\pi$ to rewrite
$A_4$ as
\eqn\amptwo{\eqalign{A_4=({\kappa^2\over 4\pi})&
\left[{\Gamma(-1+x(1+\cos\theta))
\Gamma(-1+x(1-\cos\theta))\over \Gamma(-2+2x)}\right]^2\cr
&\times\left({\sin(\pi x(1+\cos\theta))\sin(\pi x(1-\cos\theta))\over\sin
2\pi x}\right).\cr}}
{}From the Stirling approximation $\Gamma(u)\sim\sqrt{2\pi}u^{u-1/2}e^{-u}$
for large $u$, we obtain in the limit $x\to\infty$
\eqn\ampthree{\eqalign{A_4\sim&\left[{\left(x(1+\cos\theta)\right)
^{x(1+\cos\theta)}
\left(x(1-\cos\theta)\right)^{x(1-\cos\theta)}\over (2x)^{2x}}\right]^2\cr
&\times\left({\sin(\pi x(1+\cos\theta))\sin(\pi x(1-\cos\theta))\over\sin
2\pi x}\right).\cr}}
Note that the exponential terms cancel automatically. From \ampthree\ we
notice that the powers of $x$ in the first factor also cancel. $A_4$
then reduces in the limit $x\to\infty$ to
\eqn\amp{\eqalign{A_4\sim\left({1+\cos\theta\over 2}\right)^{2x(1+\cos\theta)}
&\left({1-\cos\theta\over 2}\right)^{2x(1-\cos\theta)}\cr
&\times\left({\sin(\pi x(1+\cos\theta))\sin(\pi x(1-\cos\theta))\over\sin
2\pi x}\right).\cr}}
The poles in the third factor in \amp\ are just the usual $s$-channel poles.
It follows from \amp\ that for $\theta\neq 0,\pi$~
$A_4 \to e^{-f(\theta)x}$ as $x\to\infty$,
where $f$ is some positive definite function of $\theta$.
Hence the 4-point function vanishes exponentially with the winding radius
away from the poles.

In general, for finite $R$ and fixed $v$ the strings may scatter into
longitudinally excited final states, {\it i.e.} states not satisfying
the above assumption that $m'=m$. The $4$-point amplitude for each
transition still vanishes exponentially with $R$.  A simple counting
argument shows that the total number of possible final states for a
given $R$ is bounded by a polynomial function of $R$. This counting
argument proceeds as follows:

Without loss of generality, we may assume that our incoming states have
$N=\widetilde{N}=m=0$ with fixed $R$ and $v$. We relax the assumption of
no logitudinal excitation to obtain outgoing states with nonzero $m$.
We still consider $n=1$ winding states for simplicity. Our scattering
configuration can now be described by the incoming momenta
\eqn\imom{\eqalign{p^\mu_{1R,L}&=(E,E\vec v,\pm R)\cr
p^\mu_{2R,L}&=(E,-E\vec v,\pm R).\cr}}
and the outgoing momenta
\eqn\omom{\eqalign{-p^\mu_{3R,L}&=(E_1,E_1\vec w_1,{m\over 2R}\pm R)\cr
-p^\mu_{4R,L}&=(E_2,-E_2\vec w_2,-{m\over 2R}\pm R).\cr}}
Note that in general $E_1$ and $E_2$ are not equal to $E$. Without loss
of generality, we take $m$ to be positive. From
conservation of momentum, however, we have
\eqn\conserv{\eqalign{E_1+E_2&=2E\cr E_1\vec w_1&=E_2\vec w_2.\cr}}
It follows from the energy momentum relations for the ingoing and
outgoing momenta that
\eqn\enmom{\eqalign{E^2(1-v^2)&=R^2-2\cr
E_1^2(1-w_1^2)&=2N_1-2+\left({m\over 2R}+R\right)^2\cr
E_2^2(1-w_2^2)&=2N_2-2+\left(-{m\over 2R}+R\right)^2,\cr}}
where $N_1$ and $N_2$ are the number operators for the the right movers
of the outgoing states.

Subtracting the third equation in \enmom\ from the second equation and
using \conserv\ we obtain the relation
\eqn\nme{N_1-N_2+m=(E_1-E_2)E.}
{}From the first equation in \enmom\ it follows that $E$ is bounded by
some multiple of $R$ for fixed $v$. It then follows from the first
equation in \conserv\ that both $E_1$ and $E_2$ are bounded by a
multiple of $R$. So from \nme\ we see that $N_1-N_2+m$ is bounded by
some quadratic polynomial in $R$. We now add the last two equations in
\enmom\ to obtain
\eqn\eenn{E_1^2(1-w_1^2)+E_2^2(1-w_2^2)=2N_1+2N_2+2R^2+{m^2\over
2R^2}-4.}
The left hand side of \eenn\ is clearly bounded by a quadratic
polynomial in $R$. It follows that $N_1+N_2$ is also bounded by a
quadratic polynomial, and that so are $N_1$ and $N_2$ and also, then,
$N_1-N_2$. From the boundedness of $N_1-N_2+m$ it therefore follows that
$m$ is bounded by a polynomial in $R$. Therefore
the total number of possible distinct excited states (numbered by $m$)
is bounded by a polynomial in $R$. The above argument also goes through
for the case of a nonzero initial momentum number. For each transition,
however, one can show that the Veneziano amplitude is dominated by an
exponentially vanishing function of $R$, from a calculation entirely
analogous to the zero-longitudinal excitation case worked out above.
Hence the total square amplitude of the scattering (obtained by
summing the square amplitudes of all possible transitions) is still
dominated by a factor which vanishes exponentially with the radius,
except at the poles at $\theta=0,\pi$ corresponding to forward and
backward scattering, which are physically equivalent for identical bosonic
strings. This is in agreement with the trivial
scattering found in section 3 and provides further evidence for the
identification of the solitonic string solution found in \dghrr\ with
the fundamental string.

The above argument can be repeated for any other type of string, including
the heterotic string\quartet. The kinematics differ slightly from the
tachyonic case but the $4$-point function is still dominated by an
exponentially vanishing factor in the large radius limit. Hence the
scattering is trivial, again in agreement with the result found in
section 3.

\newsec{Metric on Moduli Space in $D=4$}

In the low-velocity limit, multi-soliton solutions trace out geodesics in
the static solution manifold, with distance defined by the Manton metric on
moduli space manifold \manttwo. In the absence of a full time-dependent
solution to the equations of motion, these geodesics represent a good
approximation to the low-energy dynamics of the solitons. For BPS monopoles,
the Manton procedure was implemented by Atiyah and Hitchin
\refs{\atiyah,\atiyahbook}.

In this section we compute the Manton metric on moduli space for the
scattering of the soliton string solutions in $D=4$ although we expect that
the same result will hold for arbitrary $D \geq 4$.
We find that the metric is flat to lowest
nontrivial order in the string tension. This result implies trivial
scattering of the string solitons and is consistent with the results of the
previous two sections, and thus provides even more compelling evidence for the
identification of the string soliton with the underlying fundamental string.

We first return to the solution of \dghrr.
For $D=4$, $\phi=E$ and the metric simplifies to
\eqn\fourmet{ds^2=-dt^2+(dx^1)^2+e^{-E}d\vec x \cdot d\vec x.}

Manton's procedure may be summarized as follows. We first invert the
constraint equations of the system (Gauss' law for the case of BPS
monopoles). The corresponding time dependent field configuration does
not in general satisfy the time dependent field equations, but provides
an initial data point for the fields and their time derivatives.
Another way of saying this is that the initial motion is tangent to the
set of exact static solutions.  The resultant kinetic action obtained
by replacing the solution to the constraints in the action defines a
metric on the parameter space of static solutions. This metric defines
geodesic motion on the moduli space\manttwo.

We now assume that each string source possesses velocity
$\vec\beta_n, n=1,...,N$ in the two-dimensional transverse space $(23)$. This
will appear in the contribution
of the sigma-model source action to the equations of motion in the form
of ``moving" $\delta$-functions $\delta^{(2)}(\vec x-\vec a_n(t))$, where
$\vec a_n(t)\equiv \vec A_n + \vec\beta_n t $ (here $\vec A_n$ is the
initial position of the $n$th string source).

The equations of motion following from $S$ are complicated and nonlinear
(see \dghrr),
and it is remarkable that such a simple ansatz as section 2 could provide a
solution in the static limit. In the time dependent case, we are even less
likely to be so fortunate. In order to make headway in solving even the
$O(\beta)$ time dependent constraints, we make the simplifying assumption
that $8G\mu << 1$ (this is equivalent to assuming that each cosmic string
produces a small deficit angle). It turns out that to linear order in $\mu$
an $O(\beta)$ solution to the constraint equations is given by
\eqn\orderbeta{\eqalign{e^{-E(\vec x,t)}&=1-8G\mu\sum_{n=1}^N{\ln(
\vec x - \vec a_n(t))}\cr g_{00}&=-g_{11}=-1,\qquad g^{00}=-g^{11}=-1\cr
g_{ij}&=e^{-E}\delta_{ij},\qquad g^{ij}=e^E\delta_{ij}\cr
g_{0i}&=8G\mu\sum_{n=1}^N{\vec\beta_n\cdot \hat x_i \ln(\vec x - \vec a_n(t))},
\qquad g^{0i}=e^Eg_{0i}\cr
H_{10j}&=\partial_j e^E\cr
H_{1ij}&=\partial_i g_{0j} - \partial_j g_{0i},\cr}}
where $i,j=2,3$.

The kinetic Lagrangian is obtained by replacing the expressions for the
fields in \orderbeta\ in $S$. Since \orderbeta\ is a solution to
order $\beta$, the leading order terms in the action (after the
quasi-static part) is of order $\beta^2$. The source part of the action
($S_2$) now represents moving string sources, and its only contribution to
the kinetic Lagrangian density is of the form $(\mu/2)\beta_n^2$ for each
source. The nontrivial elements of the metric on moduli space must therefore
be read off from the $O(\beta^2)$ part of the massless fields effective action.
In principle one must add a Gibbons-Hawking surface term (GHST)
in order to cancel the double derivative terms in $S$ (see
\refs{\gh\ghp\briho\rkmant{--}\monscat}). In this case, however, the GHST
vanishes to $O(\beta^2)$. To lowest nontrivial order in $\mu$, the kinetic
Lagrangian density is computed to be
\eqn\lkin{{\cal L}_{kin}={1\over 2\kappa^2}\left(2\dot E^2-(\partial_m
g_{0k})^2\right).}
Henceforth we simplify to the case of two strings with velocities
$\vec\beta_1$ and $\vec\beta_2$ and positions $\vec a_1$ and $\vec a_2$.
Let $\vec X_n\equiv \vec x - \vec a_n, n=1,2$.
Our moduli space consists of the configuration space of the relative
separation vector $\vec a\equiv \vec a_2 - \vec a_1$. We now compute the
metric on moduli space by integrating \lkin\ over the $(23)$ space.
It turns out that the self-terms vanish on integration over the
two-space. We are then left with the interaction terms, which may be
written as
\eqn\lint{{\cal L}_{int}={64G^2\mu^2\over \kappa^2}
\left[{2(\vec\beta_1 \cdot \vec X_1)(\vec\beta_2 \cdot \vec X_2)\over
X_1^2 X_2^2} -{(\vec\beta_1 \cdot \vec\beta_2)(\vec X_1\cdot \vec X_2)\over
X_1^2 X_2^2}\right].}
The most general answer obtained by integrating \lint\
over the transverse two-space is of the form
\eqn\intform{L_{int}(\vec a)=2f(a)\vec\beta_1 \cdot \vec\beta_2 +
2g(a)(\vec\beta_1\cdot\hat a)(\vec\beta_2\cdot\hat a).}
We compute $f$ and $g$ by integrating \lint\ for only two
configurations. In both cases, $\vec\beta_1$ is parallel to $\vec\beta_2$. The
first case has the velocities parallel to $\vec a$ and yields
\eqn\para{L_{int}(a)=(2f+2g)\beta_1\beta_2}
while the second case has the velocities perpendicular to $\vec a$ and
yields
\eqn\perp{L_{int}(a)=2f\beta_1\beta_2.}
In this way we can compute both $f$ and $g$.
A slightly tedious but straightforward computation yields
\eqn\fgans{g=-2f=-{64g^2\mu^2\pi\over \kappa^2}\left(1-{\ln 2\over
2}\right),}
and thus all the metric elements are constant. In two-dimensions, this implies
that the metric on moduli space is flat (being of the form
$dr^2+Ar^2d\theta^2$, where $A$ is a constant), and therefore has
straight-line geodesics in the static solution manifold. To this
approximation, then, the low-energy scattering is trivial, i.e. the strings
do not deviate from their initial trajectories.

\newsec {Discussion}

In this section we first summarize the results obtained in this paper
and then discuss their physical implications.
In section 2 we outlined Dabholkar {\it et al.}'s multi-string soliton
solutions of the $D=10$ supergravity super Yang-Mills field equations. These
solitons resemble multi-BPS monopole solutions in that
their existence derives from a zero-force condition. Other similarities
include the saturation of a Bogomolnyi bound. The zero force condition
for parallel string solitons with the same orientation arises as a result
of the cancellation of long-range forces of exchange of the massless modes
of the string (graviton, axion and dilaton), just as the zero force condition
for equally charged monopoles results from the cancellation of
the attractive coulomb force resulting from scalar (Higgs) exchange with the
repulsive coulomb force resulting from vector (gauge) exchange. The
force cancellation was seen from the test string approximation, which
examines the force on a static test soliton in the background of the
fields produced by a source soliton.

We used the test-string approach in the next section to study
the dynamics of the string solitons. In particular, we
considered the motion of a test string in the background of a source
string. Surprisingly, we found that the velocity dependent forces vanish
as well. Were we to extrapolate this result to soliton-soliton
scattering, we would expect trivial scattering (i.e. no deviation from
initial scattering trajectories).

In section 4 we approached the string soliton scattering from a
string-theoretic (vertex operator) point of view. If the string soliton
solutions of Dabholkar {\it et al.} are to be identified with
infinitely long fundamental strings, the result of section 3
should agree with the corresponding Veneziano amplitude calculation of
fundamental string scattering.  We performed this computation for the
scattering of two macroscopic winding state strings in the large $R$
limit and once again found trivial scattering.

We returned in section 5 to the study of the dynamics of string solitons with
a computation of the Manton metric on moduli space for these solutions.
This metric describes the geodesics traced out by the multi-soliton
solutions in the static solution manifold in the low velocity limit.
A calculation of this metric in this section for the case $D=4$ to lowest
nontrivial order in the string tension yielded a flat metric, which also
implies trivial scattering.

The agreement between the results of sections 3 and 5 on the one hand
and that of section 4 provides compelling dynamical evidence for the
identification of the string soliton with the underlying fundamental string.
It is therefore likely that these solitons can be used to describe extended
string states in semi-classical string theory, in much the same way that
solitons in ordinary field theory are used to describe extended
particle states. We are especially interested in discovering inherently
``string-like" solutions, whose behaviour differs from already known
configurations in field theory and which will give us a better
understanding of string-theoretic effects in spacetime at the Planck
scale. The solutions studied here seem
to exhibit rather surprising behaviour (trivial scattering). This
dynamic behaviour differs markedly from that of the magnetic
monopole, with which the string solitons share several static features,
such as a zero static force condition and the saturation of a Bogomolnyi
bound. In this scenario, it seems that in the low-energy limit the soliton
strings also obey a zero dynamical force.

It is therefore likely that a further study of these and related
solutions (such as the fivebrane solutions in
\refs{\dfluone\dflutwo{--}\strom}\ and their exact extensions in
\refs{\rkinst\chsone{--}\chstwo}) in string theory will lead us eventually to
a better understanding of nonperturbative string theory.

\bigbreak\bigskip\bigskip\centerline{{\bf Acknowledgements}}\nobreak
I wish to thank Curtis Callan, Atish Dabholkar and Michael Duff
for helpful discussions.

\vfil\eject
\listrefs
\bye